\RequirePackage{fix-cm}
\documentclass[twocolumn,epjc3]{svjour3}  
\smartqed  % flush right qed marks, e.g. at end of proof
\RequirePackage{graphicx}
\RequirePackage{amsmath}
\RequirePackage{amssymb}
\journalname{Eur. Phys. J. A}
\begin{document}

\title{Determination of two-photon exchange via $e^+p/e^-p$ Scattering with CLAS12}


\author{Jan~C.~Bernauer
        \and
        Volker D. ~Burkert
        \and
        Ethan~Cline
        \and
        Axel~Schmidt 
        \and  
        Youri ~Sharabian 
}


\institute{Jan~C.~Bernauer \and Ethan~Cline \at Center for Frontiers in Nuclear Science, Stony Brook University, Stony Book, NY, USA
\and Jan~C.~Bernauer \at Riken BNL Research Center, Upton, NY, USA 
   \and
    Volker~Burkert \and Youri~Sharabian \at Thomas  Jefferson  National  Accelerator  Facility,  Newport  News,  Virginia, USA \label{jlab}
          \and
          Axel~Schmidt \at   The  George  Washington  University,  Washington,  DC, USA \label{gwu}         
}

\date{Received: date / Accepted: date}

\maketitle

\begin{abstract}
The proton elastic form factor ratio shows a discrepancy between measurements using the Rosenbluth technique in unpolarized beam and target experiments and measurements using polarization degrees of freedom. The proposed explanation of this discrepancy is uncorrected hard two-photon exchange (TPE), a type of radiative correction that is conventionally neglected. The effect size and agreement with theoretical predictions has been tested recently by three experiments. While the results support the existence of a small two-photon exchange effect, they cannot establish that theoretical treatments are valid. At larger momentum transfers, theory remains untested. This proposal aims to measure two-photon exchange over an extended and so far largely untested $Q^2$ and $\varepsilon$ range with high precision using the {\tt CLAS12} experiment. Such data are crucial to clearly confirm or rule out TPE as the driver for the discrepancy as well as test several theoretical approaches, believed valid in different parts of the tested $Q^2$ range.
\keywords{proton form factors \and two-photon exchange}
\end{abstract}

\section{Introduction}

\begin{figure*}[t]
\centerline{\includegraphics[width=0.8\linewidth]{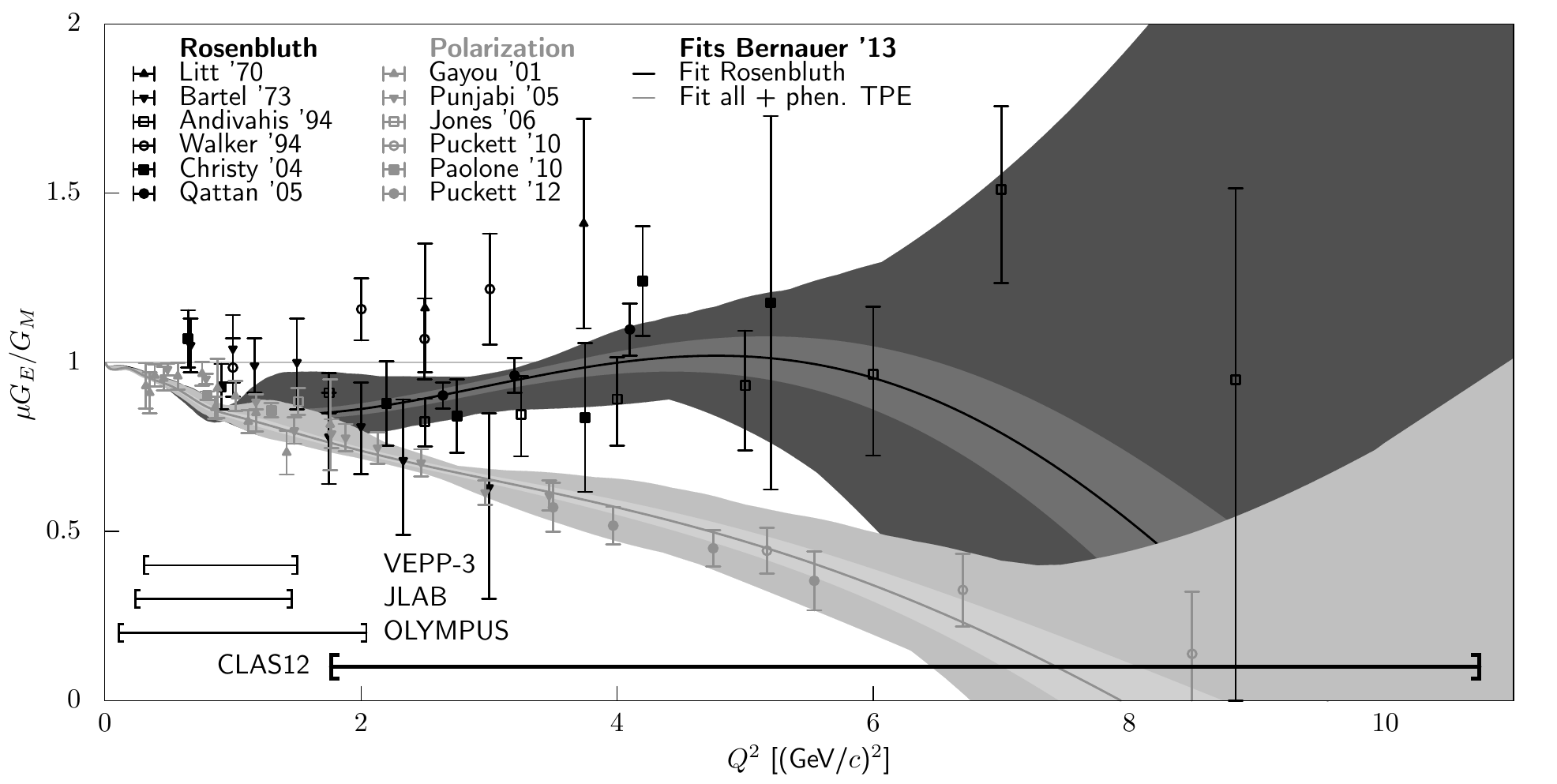}}
\caption{\label{figratio}The proton form factor ratio $\mu G_E/G_M$, as determined via Rosenbluth-type (black points, from~\cite{Lit70,Bar73,And94,Wal94,Chr04,Qat05}) and polarization-type (gray points, from~\cite{Gay01,Pun05,Jon06,Puc10,Pao10,Puc12}) experiments. While the former indicate a ratio close to 1, the latter show a distinct linear fall-off. Curves are from a phenomenological fit~\cite{Ber14}, to either the Rosenbluth-type world data set alone (dark curves) or to all data, then including a  phenomenological two-photon-exchange model. We also indicate the coverage of earlier experiments as well as of the experiment described below.}
\end{figure*}

Over more than half a century, proton elastic form factors have been extracted from electron-proton scattering experiments with unpolarized beams over a large range of four-momentum transfer squared, $Q^2$, via the so-called Rosenbluth separation. The data indicate that the form factor ratio $\mu G_E/G_M$ is in agreement with scaling, i.e., that the ratio is close to 1. This ratio is also accessible via polarized beams with fundamentally different kinematics, and, especially at large $Q^2$, improved precision. In contrast to the unpolarized result, the data indicate a roughly linear fall-off of the ratio. Some results of the different experimental methods, as well as recent fits, are compiled in Fig.~\ref{figratio}. The two data sets are clearly inconsistent with each other, indicating that one method (or both) are failing to extract the proton's true form factor ratio. The resolution of this ``form factor ratio puzzle" is crucial to advance our knowledge of the proton form factors.

The differences observed by the two methods have been attributed to two-photon exchange (TPE) effects~\cite{Gui03,Car07,Arr11,Afa17}, poised to affect especially the Rosenbluth method data. Two-photon exchange corresponds to the group of diagrams in the second order Born approximation of lepton scattering where two photon lines connect the lepton and proton. The case where one of these photons has negligible moment, the so-called ``soft'' case,  is included in the standard radiative corrections, like Ref.~\cite{Mo69,Max00}. The ``hard'' part, where both photons can carry considerable momentum, however, is not, but has been the focus of ongoing theoretical work. 

To evaluate the theoretical prescriptions and test if TPE is indeed the solution of the puzzle, precise measurements over a wide $Q^2$ range are required. The most straightforward access to TPE is via measurement of the ratio of elastic $e^+p/e^-p$ scattering, 
\begin{equation} 
R_{2\gamma} = \frac{\sigma_{e^+}}{\sigma_{e^-}} \approx 1 + 2\delta_{TPE} \, ,
\end{equation}
here, $\delta_{TPE}$ is the correction to the Born level cross section introduced by TPE.

We propose a new definitive measurement of the TPE effect that would be possible with a positron source at CEBAF. By alternately scattering positron and electron beams from a liquid hydrogen target and detecting the scattered lepton and recoiling proton in coincidence with the large acceptance CLAS-12 spectrometer, the magnitude of the TPE contribution between $Q^2$ values of 2 and 10~GeV$^2$ could be significantly constrained. With such a measurement, the question of whether or not TPE is at the heart of the ``proton form factor puzzle'' could be answered definitively.

\subsection{Previous work}

\begin{figure}[htpb]
\centerline{\includegraphics[width=\linewidth]{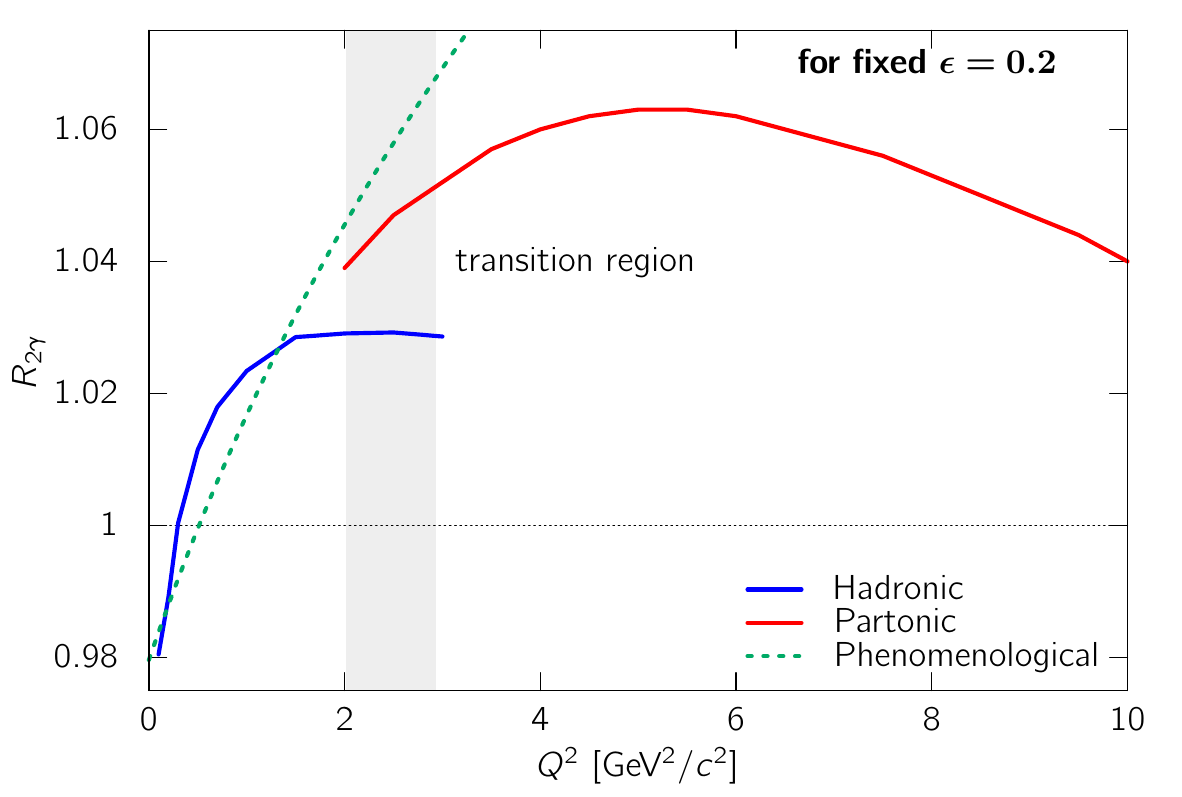}}
\caption{ A comparison of three classes of model-dependent predictions for $R_{2\gamma}$, shown as function of $Q^2$ for fixed $\epsilon=0.2$. The hadronic prediction is the $N+\Delta$ calculation from Ref.~\cite{Blu17}. The partonic prediction is the Gaussian GPD model from Ref.~\cite{Afa05}. The phenomenological prediction comes from Ref.~\cite{Ber14}, and is shown in many of the figures that follow. The validity of hadronic approaches diminishes at higher $Q^2$, while the validity of partonic approaches diminishes when $Q^2$ is small. Between $2 \lessapprox Q^2 \lessapprox 3$~GeV$^2/c^2$, it would be natural to expect some transition between the two approaches.  }
\label{fig:tpe_calc}
\end{figure}

One significant challenge is that hard TPE cannot be calculated in a model-independent way. There are several model-dependent approaches. A full description of the available theoretical calculations are outside of the scope of this letter. Suffice it to say that they can be roughly divided into two groups: hadronic calculations, e.g.\ \cite{Blu17}, which should be valid for $Q^2$ from 0 up to a couple of GeV$^2$, and GPDs based calculations, e.g.~\cite{Afa05}. The latter give a good description of nucleon form factors and wide-angle Compton scattering at JLAB kinematics and should become valid for $Q^2>1$, where the early onset of scaling is observed in DIS. At these scales, point-like quarks start to be resolved and the emissions of quarks from and re-absorption into a nucleon are described by GPDs, the overlap of light cone wave functions. A comparison of examples of these two approaches is shown in Fig.~\ref{fig:tpe_calc}.

\begin{figure}[t]
\centerline{\includegraphics[width=\linewidth]{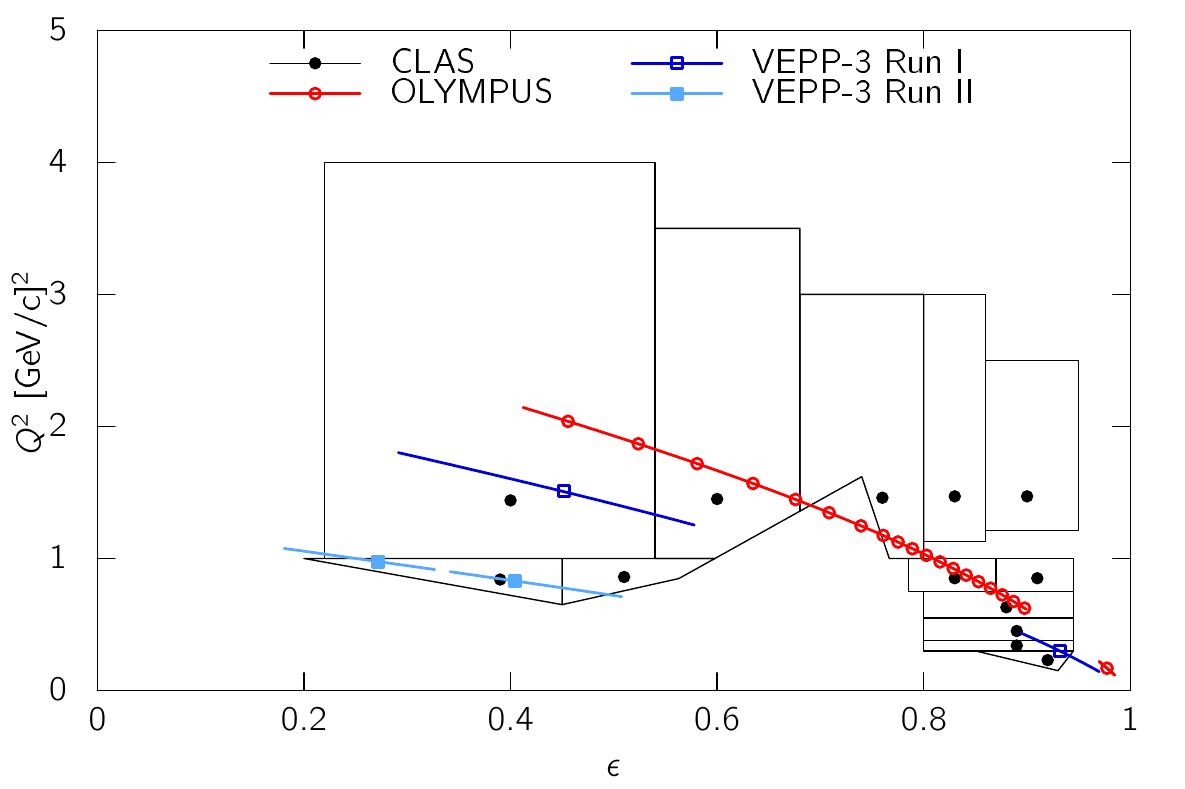}}
\caption{Kinematics covered by the three recent experiments to measure the two-photon exchange contribution to the elastic $ep$ cross section. The beam energy in the CLAS experiment was not fixed. The black polygons signify the phase space of data projected to the black (round) points.}
\label{reach}
\end{figure}

Three contemporary experiments measured the size of TPE, based at VEPP-3 \cite{Rac15}, Jefferson Lab (CLAS, \cite{Mot13,Adi15,Rim17}) and DESY (OLYMPUS, \cite{Hen17}). These experiments measured the ratio of positron-proton to electron-proton elastic cross sections. 
The kinematic reach of the three experiments is shown in Fig.~\ref{reach}. The kinematic coverage in these experiments is limited to $Q^2 < 2$~GeV$^2$, and $\varepsilon > 0.5$, where the two-photon effects are expected to be small, and systematics of the measurements must be extremely well controlled.
Comparisons of the data with theoretical predictions find overall poor agreement, an indication that TPE is not fully understood from theory. Compared to phenomenological predictions \cite{Ber14}, the agreement is good, indicating that TPE can indeed explain the majority of the discrepancy at the tested kinematics. However, at the highest $Q^2$ points, the predictions over-shoot the data considerably, pointing towards the possibility that TPE might not sufficiently explain the discrepancy at higher $Q^2$.

%Figure~\ref{figdiff} depicts the difference of the data of the three experiments to the calculation by Blunden et al.~\cite{Blu17} and the phenomenological prediction by Bernauer et al.~\cite{Ber14}. It can be seen that the three data sets are in good agreement which each other, and appear about 1\% low compared to the calculation. The prediction appears closer for most of the $Q^2$ range, however over-predicts the effect size at large $Q^2$. This is worrisome, as this coincides with the opening of the divergence in the fits depicted in Fig.~\ref{figratio} and might point to an additional effect beyond TPE that drives the difference. The combination of the experiments prefer the phenomenological prediction with a reduced $\chi^2$ of 0.68, the theoretical calculation achieves a reduced $\chi^2$ of 1.09, but is ruled out by the normalization information of both the CLAS experiment and OLYMPUS to a 99.6\% confidence level. No hard TPE is ruled out with a significantly worse reduced $\chi^2$ of 1.53.

%The current status can be summarized as such:
%\begin{itemize}
%\item TPE exists, but is small in the covered region;
%\item Hadronic theoretical calculations, supposed to be valid in this kinematical regime, might not be good enough yet;
%\item Calculations based on GPDs, valid at higher $Q^2$, are so far not tested at all by experiment;
%\item A comparison with the phenomenological extraction allows for the possibility that the discrepancy might not stem from TPE alone.
%\end{itemize}
We refer to \cite{Afa17} for a more in-depth review. The uncertainty in the resolution of the ratio puzzle jeopardizes the extraction of reliable form factor information, especially at high $Q^2$, as covered by the Jefferson Lab 12 GeV program. Clearly, new data are needed. 

\section{Proposed experiment}

Theories and phenomenological extractions predict a roughly proportional relationship of the TPE effect with $1-\varepsilon$ and a sub-linear increase with $Q^2$. However, interaction rates drop sharply with smaller $\varepsilon$ and higher $Q^2$, corresponding to higher beam energies and larger electron scattering angles. This puts the interesting kinematic region out of reach for storage-ring experiments, and handicaps external beam experiments with classic spectrometers with comparatively small acceptance. 

\begin{figure}[t!]
\begin{center}
\includegraphics[width=\linewidth]{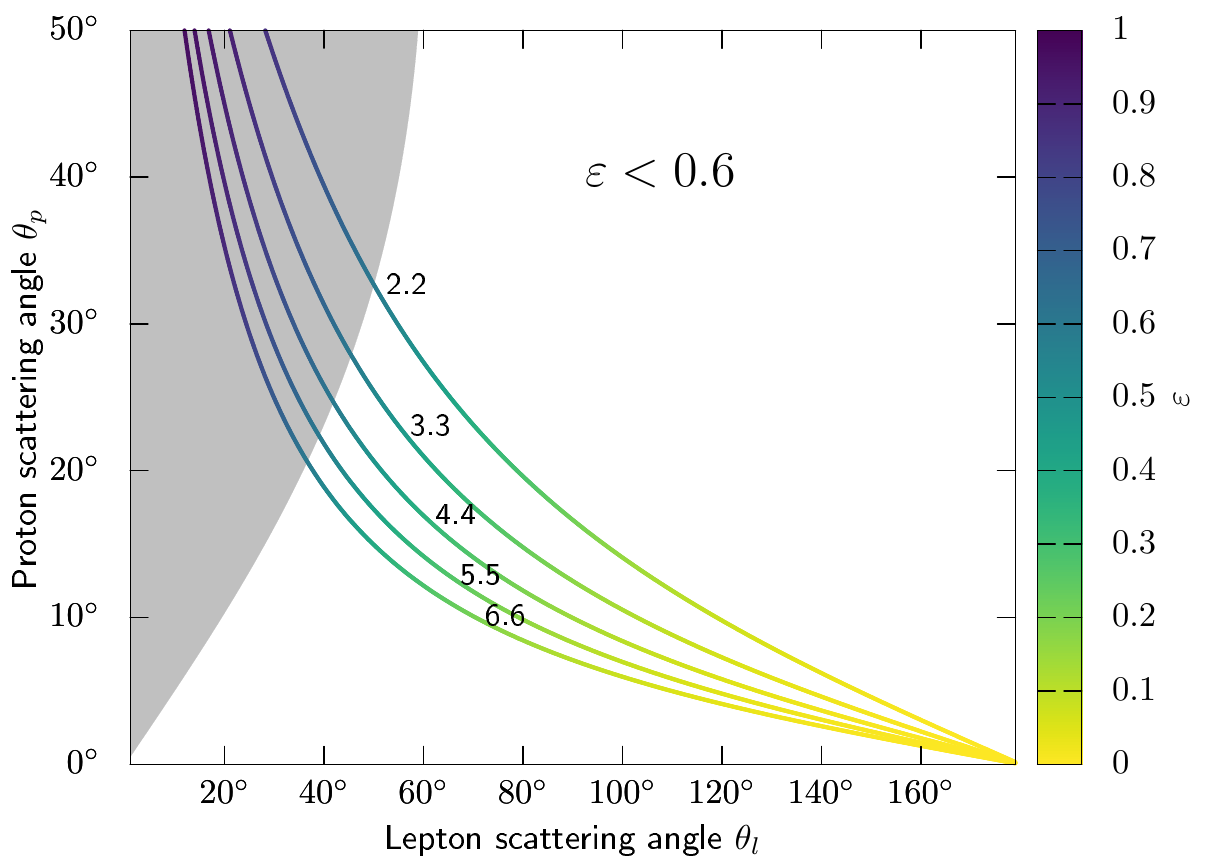}
\caption{Polar angle correlation and $\varepsilon$ coverage for lepton and proton. Each line represents a different beam energy. For the shaded area, $\epsilon>0.6$. }
\label{angle_reach}
\end{center}
\end{figure}

\begin{figure}[h]
\begin{center}
\includegraphics[width=\linewidth]{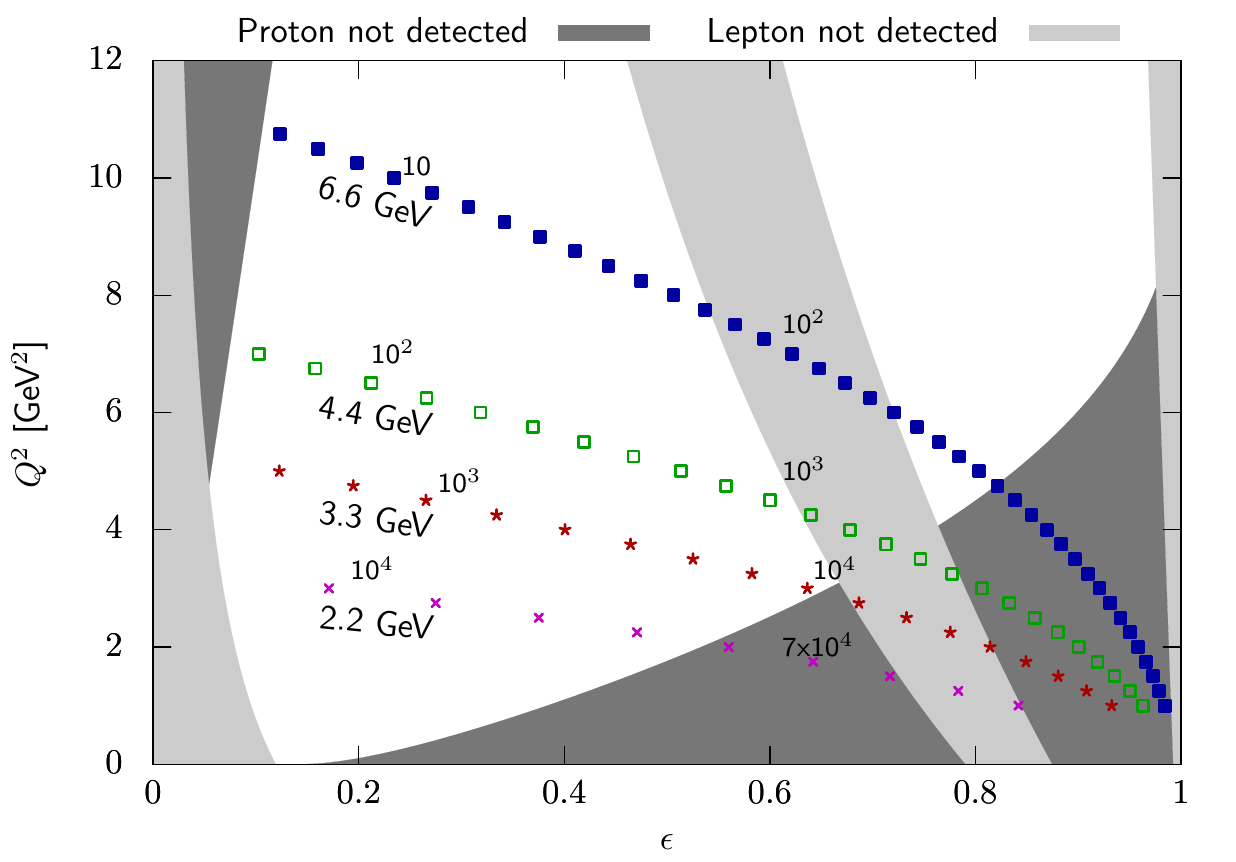}
\includegraphics[width=\linewidth]{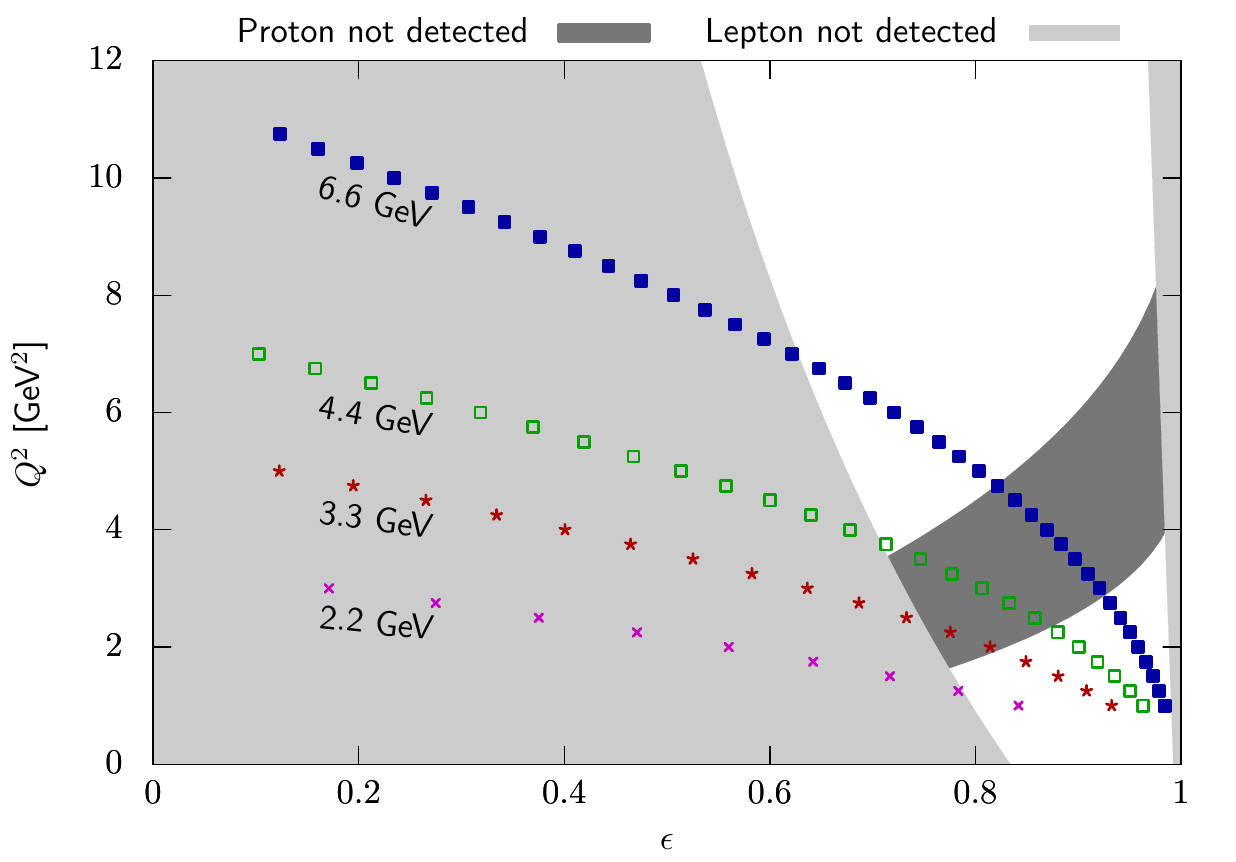}
\caption{Proposed kinematics for energies 2.2, 3.3, 4.4, 6.6 GeV in the $\varepsilon$ - $Q^2$ plane. Shaded areas are excluded by the detector acceptance. Top: proposed experiment; Bottom: standard setup. The numbers in the top plot indicate expected counts per hour.}
\label{Q2eps}
\end{center}
\end{figure}

\begin{figure}[t!]
\begin{center}
\includegraphics[width=\linewidth]{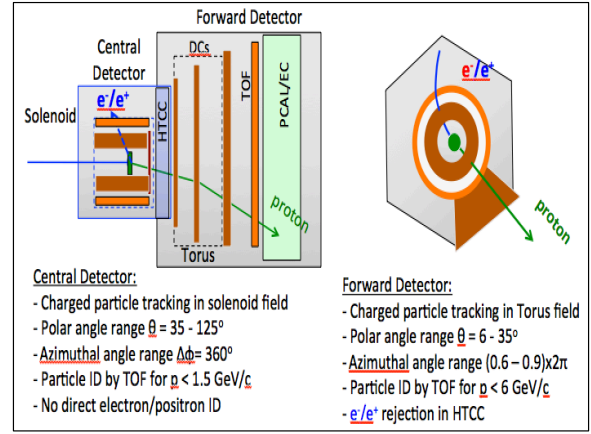}
\caption{{\tt CLAS12} configuration for the elastic $e^-p/e^+p$ scattering experiment. The central detector will detect the electron/positrons, and the bending in the solenoid magnetic field will be identical for the same kinematics. The proton will be detected in the forward detector part. The torus field direction will be the same in both cases. The deflection in $\phi$ due to the solenoid fringe field will be of same in magnitude of $\Delta\phi$ but opposite in direction. The systematic of this shift can be controlled by doing the same experiment with reversed solenoid field direction.}
\label{2gamma-exp}
\end{center}
\end{figure}

With the large acceptance of {\tt CLAS12}, combined with an almost ideal coverage of the kinematics, measurements of TPE across a wide kinematic range are possible, complementing the precision form factor program of Jefferson Lab, and testing both hadronic (valid at the low $Q^2$ end)  as well as GPD-based (valid at the high $Q^2$-end) theoretical approaches. Figure~\ref{angle_reach} shows the angle correlation between the lepton and the proton for different beam momenta. There is a one-to-one correlation between the lepton scattering angle and the proton recoil angle. For the kinematics of interest, say $\varepsilon < 0.6$ and $Q^2 > 2$~GeV$^2$ for the chosen beam energies from 2.2 to 6.6 GeV, 
nearly all of the lepton scattering angles falls into a polar angle range from $40^\circ$  to $125^\circ$, and corresponding to the proton polar angle range from $8^\circ$ to $35^\circ$. These kinematics are most suitable for accessing the TPE contributions. The setup will also be able to measure the reversed kinematics with the electrons at forward angle and the protons at large polar angles, i.e.\ the standard {\tt CLAS12} configuration of DVCS and most other experiments. While the two-photon exchange is expected to be small in this range, the sign change in TPE seen in the experiments, but not predicted by current theories, can be studied. \newline
Figure~\ref{Q2eps} shows the expected elastic scattering rates  covering the ranges of highest interest, with $\varepsilon < 0.6$ and $Q^2 = 2 - 10$~GeV$^2$. Sufficiently high statistics can be achieved within 10 hrs for the lowest energy and within 1000 hrs for the highest energy, to cover the full range in kinematics. Note that all kinematic bins will be measured simultaneously at a given energy, and the shown rates are for the individual bins in ($Q^2$, $\varepsilon$) phase-space.
In order to achieve the desired kinematics reach in $Q^2$ and $\varepsilon$ the {\tt CLAS12} detection system has to be used with reversed detection capabilities for leptons. The main modification will involve replacing the current Central Neutron Detector with a central electromagnetic calorimeter (CEC), a concept that has already been studied for the {\tt CLAS12} program. The CEC will not need very good resolution, which is provided by the tracking detectors, but will only be used for trigger purposes and for electron/pion separation. The strict kinematic correlation of the scattered electron and the recoil proton will be sufficient to select the elastic events. The {\tt CLAS12} configuration suitable for this experiment is shown in Fig.~\ref{2gamma-exp}.

\subsection{Central Electromagnetic Calorimeter}

The Central Electromagnetic Calorimeter (CEC) will be used for trigger purposes to detect electrons elastically scattered under large angles and for the separation of electrons (positrons) and charged pions. The CEC will be a new detector for {\tt CLAS12}, but based on a mature design from the original {\tt CLAS12} concept. The detector will use a novel tungsten powder/scintillating fiber calorimetry technology first proposed in 1999. This original proposal was to develop a compact, high-density fast calorimeter with good energy resolution at polar angles greater than 35$^\circ$ for the {\tt CLAS12} spectrometer ~\cite{Car01}, and occupy the radial space of $\approx$ 10 cm to fit inside the Central Solenoid. The tungsten powder technology has the benefits of compactness, homogeneity, simplicity, and favorable readout characteristics.

For the proposed elastic scattering experiment, the CEC would replace the current Central Neutron Detector, which occupies approximately the same radial space and polar angle range. The calorimeter will need to cover polar angles in a range of 40$^\circ$ to 130$^\circ$, and the full 2$\pi$ range in azimuth. From the original proposal there exists a prototype calorimeter that was designed, built, and cosmic-ray tested, and we plan to construct a CEC with parameters close the existing prototype calorimeter. The dimensions of the prototype's active volume are approximately length $\times$ width $\times$ height = 0.1$\times$0.1$\times$0.07~m$^3$ in volume and with 5,488 fibers (Bicron BCF-12) with 0.75~mm diameter, uniformly distributed inside the tungsten powder volume. These fibers make up 35\% of the volume and the tungsten powder is filled into the remaining volume. The final density of the tungsten powder radiator is 12~g/cm$^3$, or about 5\% higher as compared with the density of bulk lead. The overall total density of the prototype active volume is $\approx$~8.0~g/cm$^3$. The estimated signal strength is  about 75 photoelectrons per MeV. There is the possibility of increasing the density of the radiator to $\approx$~10.5~g/cm$^3$, which will lead to an increase of the detector absorption power. Also an additional increase can be obtained by simply decreasing sampling ratio, since having higher energy resolution is not a critical requirement. It has to be mentioned that due to the cylindrical shape of the CEC there will be no {\it side wall} effects. 

\subsection{Projected measurements}

\begin{figure}[t!]
\centerline{\includegraphics[width=\linewidth]{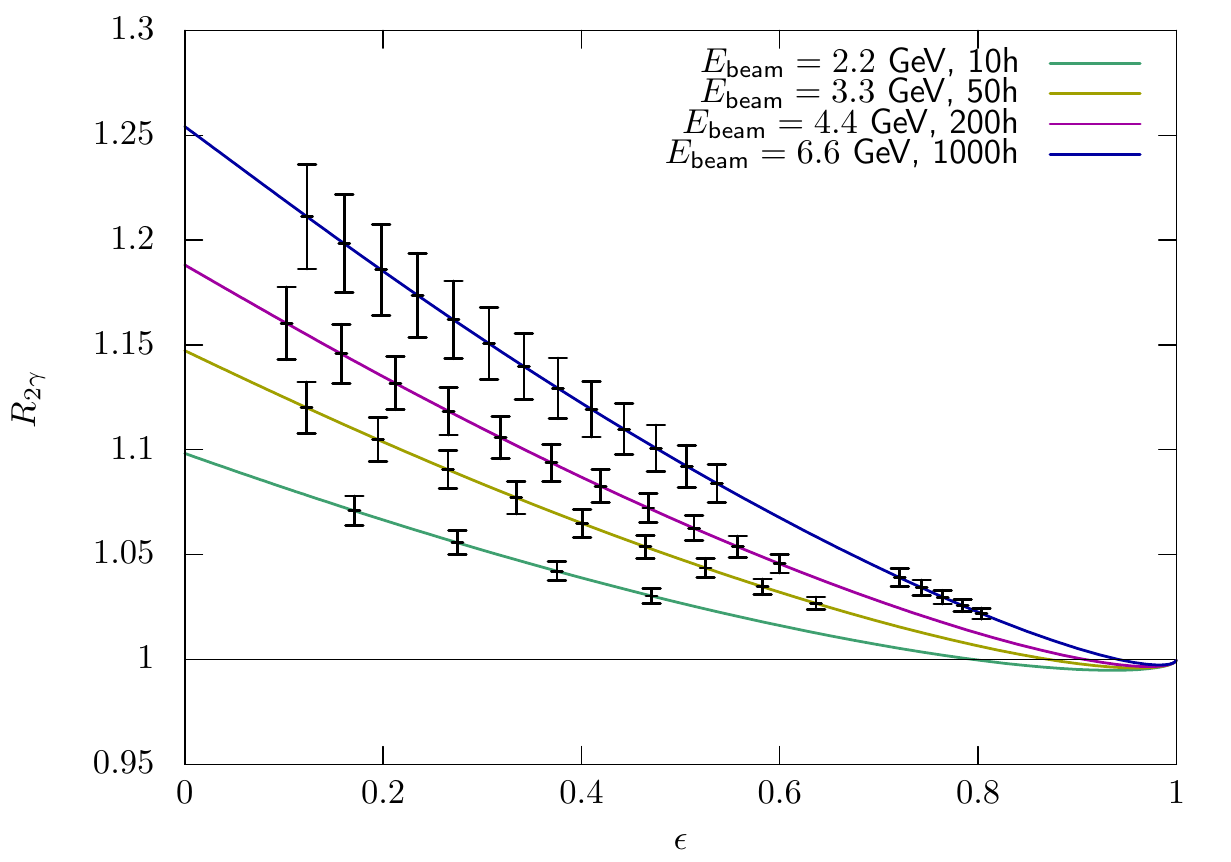}}
\caption{\label{figerrclas}Predicted effect size and estimated errors for the proposed measurement program at {\tt CLAS12}. We assume bins of constant $\Delta Q^2$=$0.25$~{GeV}$^2$. The prediction is based on \cite{Ber14}.}
\end{figure}

For the rate estimates and the kinematical coverage we have made a number of assumptions that are not overly stringent: 
\renewcommand{\labelenumi}{\it\roman{enumi})}
\begin{enumerate}
\item Positron beam currents (unpolarized), $I_{e^+} \approx 60$~nA;  
\item Beam profile, $\sigma_x,~\sigma_y < 0.4$~mm;  
\item Polarization not required, so phase space at the source can be optimized for yield and beam parameters;
\item Operate experiment with 5~cm liquid H$_2$ target and luminosity of $0.8 \times 10^{35}$~cm$^{-2} \cdot$sec$^{-1}$;
\item Use the {\tt CLAS12} Central Detector for lepton ($e^+/e^-$) detection at $\Theta_l$=$40^\circ$-$125^\circ$;
\item Use {\tt CLAS12} Forward Detector for proton detection at $\Theta_p$=$7^\circ$-$35^\circ$.
\end{enumerate}   
We propose to take data at beam energies of 2.2, 3.3, 4.4 and 6.6 GeV, for 10~h, 50~h, 200~h and 1000~h respectively, split 1:1 in electron and positron running. The expected statistical errors, together with the expected effect size (phenomenological extraction from~\cite{Ber14}) are shown in Fig.~\ref{figerrclas}. The quality of the measured data will quantify hard two-photon-exchange over the whole region of precisely measured and to-be-measured cross section data, enabling a model-free extraction of the form factors from those. It will test if TPE can reconcile the form factor ratio data where the discrepancy is most significantly seen, and test, for the first time, GPD-based calculations. 

\subsection{Systematics of the comparison between electron and positron measurements}

The main benefit to measure both lepton species in the same setup closely together in time is the cancellation of many systematics which would affect the result if data of a new positron scattering measurement is compared to existing electron scattering data. For example, one can put tighter limits on the change of detector efficiency and acceptance changes between the two measurements if they are close in time, or optimally, interleaved. \newline
For the ratio, only relative effects between the species types are relevant; the absolute luminosity, detector efficiency, etc.\ cancel. Compared to classic small acceptance spectrometers, even the requirements on the relative luminosity determination are somewhat relaxed, as all data points of one species share the same luminosity, that is, even without any knowledge of the relative normalization between species, the evolution of TPE as a function of $\varepsilon$ for constant beam momenta could be extracted. To achieve then an absolute normalization of the ratio, the relative luminosity must be controlled.  

%Precise relative measurement methods, for example based on M\o ller scattering, exist, but only work when the species is not changed. Switching to Bhabha scattering for the positron case and comparing with M\o ller scattering is essentially as challenging as an absolute measurement. More suitable is a measurement of the lepton-proton cross section itself at extreme forward angles, i.e., $\varepsilon\approx1$, where TPE should be negligible and the cross section is the same for both species.  To make use of these cancellations, it is paramount that the species switch-over can happen in a reasonable short time frame ($<1$~day) to keep the accelerator and detector setup stable. For the higher beam energies, where the measurement time is longer, it would be ideal if the species could be switched several times during the data taking period. To keep the beam properties as similar as possible, the electron beam should not be generated by the usual high quality source, but employ the same process as the positrons. 

The primary means of normalization for low current experiments in Hall B is the totally absorbing Faraday cup (FC) in the Hall B beam line. The absolute accuracy of the FC is better than 0.5\% for currents of 5~nA or greater. The FC can be used in $e^+$/$e^-$ beams with up to 500 W, which should not be a limitation for experiments in Hall B with {\tt CLAS12}. The relative accuracy for the ratio of electrons to positrons should be at least as good as the absolute accuracy. The only known difference between electrons and positrons is the interaction of $e^+$ and $e^-$ with the vacuum window at the entrance to the FC, which is a source of M\o ller  scattering for electrons and a source of Bhabha scattering for positrons. The FC design contains a strong permanent magnet inside the vacuum volume and just after the window. This magnet is meant to trap (most of) the low-energy M\o ller electrons to avoid over-counting the electric charge. It will also trap (most of) the Bhabha scattered electrons from the positron beam to avoid under-counting (for positrons) the electric charge. However, there may be a remaining, likely small charge asymmetry for M\o ller and Bhabha scattered electrons in the response of the FC to the different charged beams. This effect will be studied in detail with a GEANT4 simulation. In any case, they relative efficiency of the FC can be calibrated with a measurement of $R$ at small scattering angles, i.e.\  $\varepsilon\rightarrow 1$, where TPE effects become negligible. This calibration could be performed with the Forward Tagger Calorimeter which covers down to 2.5 degrees. The high counting rates make this a simple and fast calibration.

\subsection{Radiative corrections}
For an extraction of the hard part of the two-photon exchange, the measured raw ratio has to be corrected for radiative effects, including other charge-odd contributions. These include the soft two-photon exchange, but also the interference terms from radiation off the lepton and proton. Current radiative generators, for example ESEPP \cite{Gramolin:2014pva}, or those from the A1 \cite{Ber14} and OLYMPUS experiments \cite{Hen17} allow us to include the radiative corrections as part of a full simulation, instead of a post-hoc correction factor. 

The absolute size of the correction depend strongly on the cuts applied to select elastic reactions. Here, wider cuts lead to smaller corrections, however, not necessarily to smaller uncertainties, as the wider cuts accept kinematics further away from the elastic case captured in the theoretical calculations.  

\begin{figure}[t]
\centerline{\includegraphics[width=\linewidth]{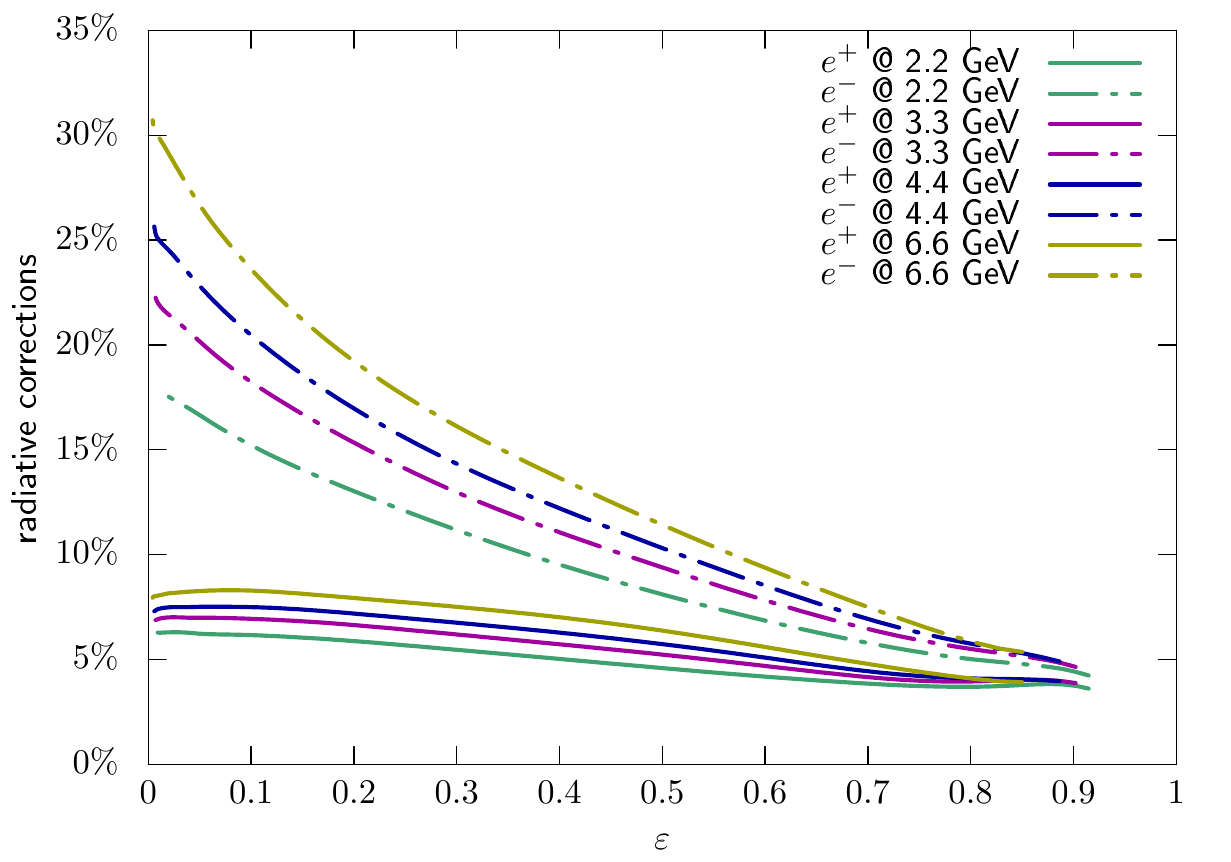}}
\caption{Estimate of the radiative corrections for the four beam energies and both lepton charges. For positrons, the charge-odd corrections have an opposite sign from the charge-even corrections, reducing the overall size and drastically reducing the $\epsilon$ dependence.}
\label{fig:radcorr}
\end{figure}

Figure~\ref{fig:radcorr} show an estimate of the radiative corrections (as corrections to a a Born level calculation) for the four beam energies and both species. Here, selection cuts are chosen to accept missing energies (i.e., energies of the radiated photon) up to 20\% of the outgoing lepton energy. Further, a 50~mrad-wide cut is applied on the lepton-angle vs.~proton-angle correlation. For positrons, the charge-odd corrections reduce the size of the overall correction, however, the correction will have the same uncertainty as for the electron case, in which the charge-odd corrections have the same sign as the charge-even part. 

\subsection{Charge-averaged cross sections}
In addition to a measurement of the ratio, the data set also allows a determination of the charge-averaged cross sections, similar to the recent results released by OLYMPUS \cite{Ber20}, however with the added benefit of good absolute normalization. While the ratio is only sensitive to the charge-odd corrections, the charge-averaged cross sections cancel these, and are only sensitive to charge-even corrections. With the data set alone, Rosenbluth separations at several $Q^2$ up to about 7 $(\mathrm{GeV}/c)^2$ will be possible and would allow a direct comparison with polarized measurements. Additionally, the data set will be a helpful addition to the world data set and will reduce fit uncertainties in the covered $Q^2$ range.

\section{Summary}
Despite recent measurements of the $e^+p/e^-p$ cross section ratio, the proton's form factor discrepancy has not been conclusively resolved, and new measurements at higher momentum transfer are needed. {\tt CLAS12}, in combination with a positron beam at CEBAF, would be the definitive measurement of TPE over a wide and highly significant kinematic range. Only one major detector configuration change would be necessary to support such a measurement, the installation of the central electromagnetic calorimeter. In designing the JLab positron source, it will be crucial for this and several other experiments to keep to a minimum the time necessary to switch between electron and positron modes, in order to reduce systematic effects.

\begin{acknowledgements}
This work was supported in part by the National Science Foundation grant number 2012114.

\end{acknowledgements}



\end{document}